\documentclass[twocolumn,showpacs,preprintnumbers,amsmath,amssymb]{revtex4}
\usepackage{graphicx}% Include figure files
\usepackage{dcolumn}% Align table columns on decimal point
\usepackage{bm}% bold math
\usepackage{epsfig}
\newcommand{\be}{\begin{equation}}
\newcommand{\ee}{\end{equation}}
\newcommand{\ba}{\begin{eqnarray}}
\newcommand{\ea}{\end{eqnarray}}
\newcommand{\bi}{\begin{itemize}}
\newcommand{\ei}{\end{itemize}}

\newcommand{\<}{\langle}
\renewcommand{\>}{\rangle}
\newcommand{\eq}{Eq.~}

\newcommand{\fig}{Fig.~}

\newcommand{\la}{\label}

\newcommand{\Nt}{N_{\tau}}

\begin{document}

% \preprint{MIT-CTP ????}

\title{Density, short-range order and the quark-gluon plasma} 

\author{Harvey~B.~Meyer}
\email{meyerh@mit.edu}
\affiliation{Center for Theoretical Physics\\ 
       Massachusetts Institute of Technology\\
                 Cambridge, MA 02139, U.S.A.}

\date{\today}

\begin{abstract}
We study the thermal part of the energy density spatial correlator 
in the quark-gluon plasma. We describe its qualitative form
at high temperatures. We then calculate it out to distances
$\approx 1.5/T$ in SU(3) gauge theory lattice simulations for
the range of temperatures $0.9\leq T/T_c\leq 2.2$. The vacuum-subtracted
correlator exhibits non-monotonic behavior, and is almost conformal by $2T_c$.
Its broad maximum at $r\approx 0.6/T$ suggests a dense medium with only
weak short-range order, similar to a non-relativistic fluid near the 
liquid-gas phase transition, where $\eta/s$ is minimal.
\end{abstract}

\maketitle

%%%%%%%%%%%%%%%
%\section{Introduction\la{sec:intro}}
%%%%%%%%%%%%%%%
Hydrodynamics calculations~\cite{huovinen} successfully described
the pattern of produced particles in heavy ion collisions 
at RHIC~\cite{Arsene:2004fa}. This early agreement between
ideal hydrodynamics and experiment has been refined in recent times.
On the theory side, the dissipative effects of shear viscosity $\eta$
have been included  in full 3d hydrodynamics calculations~\cite{paulrom,Dusling:2007gi,Song:2007fn}
and the sensitivity to initial conditions quantitatively estimated~\cite{Luzum:2008cw}
for the first time.
On the experimental side, the elliptic flow observable $v_2$,
which is sensitive to the value of $\eta$ in units of entropy density $s$, 
is now corrected for non-medium-generated two-particle correlations~\cite{:2008ed}.
The conclusion that $\eta/s$ must be much smaller than unity
has so far withstood these refinements of heavy-ion phenomenology~\cite{Luzum:2008cw}.

The smallness of $\eta/s$ was turned into the statement that the 
quark-gluon plasma (QGP) formed at RHIC is the ``most perfect liquid 
known in nature''~\cite{Hirano:2005wx}. A general question then comes to mind:
what observable can be used to characterize the liquid 
nature of a system described by a quantum field theory~\cite{Thoma:2005ym}?
And secondly, what is the QCD prediction for that observable?
This leads us to remind ourselves what the defining property 
of an ordinary liquid is. Surely the everyday-life notion
that a liquid ``has a definite volume, but no definite shape''
is inadequate in the present context.

% The spatial density-density correlator
The two-body density distribution $\rho({\bf r_1,\bf r_2})=g(r)\rho^2$ 
of an ordinary substance (such as water) of density $\rho$
behaves qualitatively differently 
in the solid, liquid  and gas phase~(see for instance~\cite{water}).
The radial distribution function $g(r)$ characterizes
the average density of particles at distance $r$ from
an arbitrarilty chosen particle.
In a dilute gas,  $g(r)$ is essentially 
equal to $1$ for $r$  greater than the size 
of a molecule. In a liquid on the other hand, 
$g(r)$ vanishes at small $r$, a reflexion
of the short-distance repulsion between molecules.
The function then rises and typically exhibits several 
gradually damped oscillations around unity.
This reflects the ``short-range order'' in the fluid, 
namely the coherent motion of closely packed molecules
up to distances a few times the molecule size.
Over longer distances, this ordering is lost. Only a perfect
crystal at low temperatures exihibits truly long-range order.
% A similar analogy with ordinary liquids 
% was already put forward by M.~Thoma~\cite{Thoma:2005ym}. 
% The author focused on the quark number density operator,
% which he calculated at leading order in the HTL framework.
% The subtraction of the $T=0$ part was not considered.

In quantum field theory, particle number is not (necessarily)
conserved, so it is not immediately clear which spatial
correlator is the closest analogue of the 
two-body density distribution in non-relativistic systems.
In QCD, the only conserved quantities are energy, momentum, 
the quark numbers and the (non-singlet) axial charges. 
The energy density is an order parameter in the theoretical limit of 
a large number of colors $N_c$, which puts it in natural
correspondence with the density $\rho$ of a non-relativistic fluid.
Hence our strategy to study the spatial correlator of 
the energy density. 

A  peculiarity of quantum field theory is 
that energy density correlations are present even when 
the average energy density is zero, i.e. at $T=0$ % zero temperature
when the partition function is saturated by the quantum vacuum.
This correlation has to be strong at short distance
in an asymptotically free theory,  $\<T_{00}(0)T_{00}({\bf r})\>\sim
r^{-8}$, on dimensional grounds. Based on the K\"allen-Lehmann 
representation, it is monotonically decreasing in $r$ at any $T$.
% (more generally, its $n^{\rm th}$ $r$-derivative has 
% sign $(-1)^n$ at any temperature).

Therefore, in order to isolate the thermal effects, we shall
consider the subtracted correlator
\be
% &&  \!\!\!\!\!\!\!\! 
\! 
G_{ee}(T,r)= \<T_{00}(0)T_{00}(0,{\bf r})\>_T - 
             \<T_{00}(0)T_{00}(0,{\bf r})\>_{T=0},
\la{eq:G} 
% \\ &&\!\!\!\! = \<T_{00}(0)T_{00}(0,{\bf r})\>^{\rm conn}_T 
% - \<T_{00}(0)T_{00}(0,{\bf r})\>^{\rm conn}_{T=0} +e^2(T).
% \nonumber
\ee
% and treat it as a relativistic analogue of the
% two-body density distribution of non-relativistic systems
% We have rewritten the expressions  in terms of connected correlators.

In the Euclidean SU($N_c$) gauge theory, the energy density operator,
\ba
T_{00} &=& \theta_{00}+\frac{1}{4} \theta\,,~~
\theta =\frac{\beta(g)}{2g}F_{\mu\nu}^aF_{\mu\nu}^a,~~
\<\theta\> = e-3p,
\\
\theta_{00} &=& \frac{1}{4} F_{ij}^aF_{ij}^a-\frac{1}{2}F^a_{0i}F^a_{0i}\,,
~~~
\<\theta_{00}\> = \frac{3}{4}(e+p)= \frac{3}{4}T\,s \,,\nonumber 
% \\ \nonumber
\ea
is composed of two terms which are separately scale-independent operators.
Here $e$ is the energy density, $p$ the pressure and 
$\beta(g)=-bg^3+\dots$ the beta function, with $b=\frac{11N_c}{3(4\pi)^2}$.
The expectation value of the `naive' energy operator $\theta_{00} $ 
is proportional to the entropy density, while the expectation value 
of the trace anomaly directly measures the deviation from the 
conformal limit where $e=3p$ (we implicitly add a constant to $T_{00}$ 
such that $\<T_{00}\>$  vanishes in the vacuum). We can thus investigate separately 
the $G_{ss}$ and $G_{\theta\theta}$ correlators, defined by replacing 
$T_{00}$ respectively by  $\frac{4}{3}\theta_{00}$ and $\theta$ in \eq\ref{eq:G}.

\section*{High-temperature behavior}
The one-loop expression for the two correlators 
in $D$ space-time dimensions is 
\ba
&& \!\!\!\!\! \<\theta_{00}(0)\theta_{00}(x)\>_{\rm 1L} = 
(8\pi b\alpha_s)^{-2}\<\theta(0)\theta(x)\>_{\rm 1L}  =   \la{eq:tl}\\
&& \!\!\!\!\! \frac{d_A}{4\pi^4}\sum_{m,n\in {\bf Z}}
\left(D-8 +16 \frac{(x_{[m]}\cdot x_{[n]})^2} {x_{[m]}^2x_{[n]}^2}\right)
\frac{1}{(x_{[m]}^2 \,x_{[n]}^2)^2 }
\nonumber
\ea
where $d_A=N_c^2-1$ and $x_{[n]} = (\frac{n}{T}+x_0,{\bf x})$, while 
$\<\theta_{00}(0)\theta(0,{\bf r})\>_{\rm 1L}$ vanishes identically.
The $m=n=0$ term gives the zero-temperature expression.

Let us consider the high-temperature regime, $T\gg g^2(T)T\gg T_c$.
In that case, the behavior of $G_{ee}$  is characterized by 
three regimes:
\begin{description}
\item[$Tr\ll g^{-2}(T)$:] both the zero and high temperature
correlators are well described by the one-loop formula, up to 
small radiative corrections. For $r\to0$, $G_{ee}(T,r)\sim -\frac{2T^4}{45r^4}$;
for $Tr>0.568$,  $G_{ee}>e^2(T)$ (see \fig\ref{fig:ee-nt6}).
\item[$g^{-2}(T)\ll Tr \ll \frac{T}{T_c}$:]
the zero temperature correlator is still described by 
perturbation theory, $\frac{3d_A}{\pi^4{r^8}}$, but 
 the high-$T$ correlator is exponentially screened, 
so necessarily  $G_{ee}<e^2(T)$.
\item[$Tr\gg \frac{T}{T_c}$:] both the zero and high temperature
correlators are exponentially screened. In the former case, the 
relevant mass  is $M_{4}\simeq 5.3 T_c$~\cite{thesis,teper}, 
corresponding to the lightest scalar glueball mass in $D=4$, 
while at high temperatures, dimensional
reduction takes places, and the screening mass is 
$(M_3/g_3^2)\cdot g^2(T)T$ with $M_3/g_3^2\simeq 2.4$ ~\cite{teper2+1}.
It is therefore clear that screening of the energy density 
is stronger at high temperatures, and hence $G_{ee}$ asymptotically approaches
$e^2(T)$ from below at a rate $e^{-M_4r}$.
\end{description}
In summary, at high temperatures $G_{ee}(T,r)-e^2(T)$
vanishes at least at two finite distances $r$:
the first time at $Tr\approx 0.568$, and the second time at 
$Tr={\rm O}(g^{-2}(T))$. 
To investigate the function $G_{ee}(T,r)$ 
at temperatures accessible in heavy-ion colliders,
we perform lattice simulations in the region $0.9T_c<T<2T_c$.

% It is probable that at temperatures accessible in heavy-ion colliders, 
% the first regime goes over directly into the third regime.
% Finally we remark that the third regime is absent in a conformal field theory such 
% as the ${\cal N}=4$ SYM theory.

\section*{Operator product expansion}
From expression (\ref{eq:tl}) and from known results~\cite{svz},
we can obtain the leading terms in the 
operator-product expansion (OPE) of these correlators,
\ba
&& \!\!\!\! \!\!\!\! 
\<\theta(0)\theta(x)\> \sim
\frac{(8\pi b\alpha_s)^2 3d_A}{\pi^4{r^8}}
-\frac{64b^2}{3}\alpha_s^2\frac{\<\theta_{00}\>}{r^4}
-32b^2\alpha_s^2\frac{\<\theta\>}{r^4} 
 \nonumber\\
% + {\rm O}(r^{-2})\,, \\
&&\!\!\!\!\!\!\!\!    \<\theta_{00}(0)\theta_{00}(x)\> \sim
\frac{3d_A}{\pi^4{r^8}} -\frac{1}{3\pi^2} \frac{\<\theta_{00}\>}{r^4}
+ {\rm O}(\alpha_s^0) \frac{\<\theta\>}{r^4}\,,
\la{eq:OPE}
% + {\rm O}(r^{-2})\,.
% \nonumber \\
% \<\theta_{00}(0)\theta(0,{\bf r})\> &\sim&
% \frac{{\rm O}(\alpha_s^2)}{r^8}
\ea
where $x=(0,{\bf r})$ and $r^{-2}$ terms and softer have been omitted.
The Wilson coefficients of the operators ${\bf 1}$, $\theta_{00}$ 
and $\theta$ are a least of ${\rm O}(\alpha_s^2)$ for the product 
$\theta_{00}(0)\theta(0,{\bf r})$.
The usefulness of the OPE in this context arises because of the exact
cancellation of the $r^{-8}$ term in the difference between finite $T$ and 
$T=0$ correlators. Using \eq\ref{eq:OPE}, we obtain the short-distance 
behavior
\be
G(T,{\bf r})
\sim -\frac{e+p}{(2\pi r^2)^2} + {\rm O}(\alpha_s^0) \frac{e-3p}{r^4}+
{\rm O}(\alpha_s r^{-4}, r^{-2})
\ee
%%%%%%%%%%%%%%%%%%%%%%%%%%%%%%%%%%%%%%%%%%%%%%%%%%%%%%%%%%%%%%%%%%%%%%%
\begin{figure}[t]
\centerline{\psfig{file=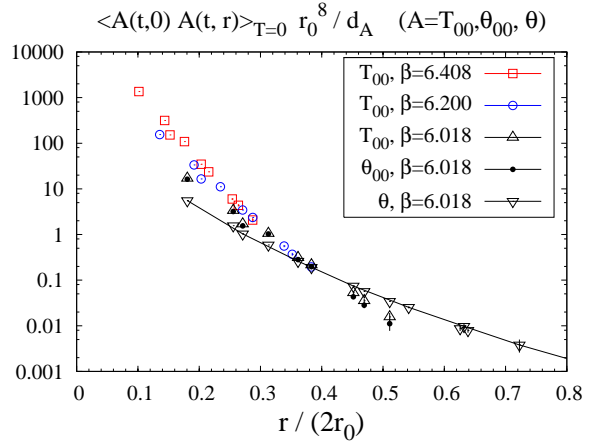,angle=-90,width=8.5cm}}
\caption{The correlators at $T=0$ at different lattice spacings
($r_0\approx 0.5$fm~\cite{necco-sommer}; the line is to guide the eye).}
\la{fig:T0}
\end{figure}
%%%%%%%%%%%%%%%%%%%%%%%%%%%%%%%%%%%%%%%%%%%%%%%%%%%%%%%%%%%%%%%%%%%%%%%
% \begin{figure}[t]
% \centerline{\psfig{file=cutoff-1p24.ps,angle=-90,width=8.5cm}}
% \caption{The energy-energy correlator at $1.24T_c$, 
% on a $6\times22^3$ lattice (open symbols) and on a $8\times28^3$ lattice 
% (closed symbols).}
% \la{fig:cutoff-1p24}
% \end{figure}
%%%%%%%%%%%%%%%%%%%%%%%%%%%%%%%%%%%%%%%%%%%%%%%%%%%%%%%%%%%%%%%%%%%%%%%
\begin{figure}[t]
\centerline{\psfig{file=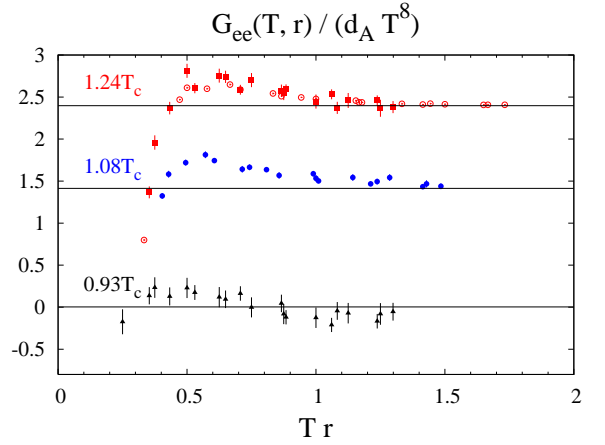,angle=-90,width=8.5cm}}
\caption{$G_{ee}$ (\eq\ref{eq:G}) across the deconfining 
phase transition, at $\beta=6.018$. At $T=1.24T_c$ we check for 
discretization errors by also showing data from $\beta=6.200$ 
(filled squares).}
\la{fig:ee-b6p018}
\end{figure}
%%%%%%%%%%%%%%%%%%%%%%%%%%%%%%%%%%%%%%%%%%%%%%%%%%%%%%%%%%%%%%%%%%%%%%%
\begin{figure}[t]
\centerline{\psfig{file=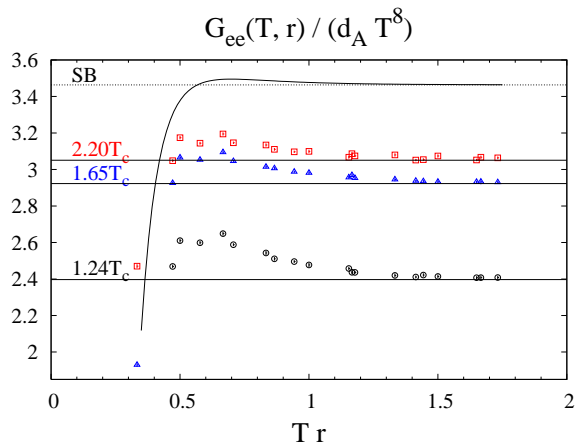,angle=-90,width=8.5cm}}
\caption{$G_{ee}$ at three temperatures,
with  $\Nt=6$. The curve corresponds to the 
correlator in the Stefan-Boltzmann limit $T\to\infty$.}
\la{fig:ee-nt6}
\end{figure}
%%%%%%%%%%%%%%%%%%%%%%%%%%%%%%%%%%%%%%%%%%%%%%%%%%%%%%%%%%%%%%%%%%%%%%%
\begin{figure}[t]
\centerline{\psfig{file=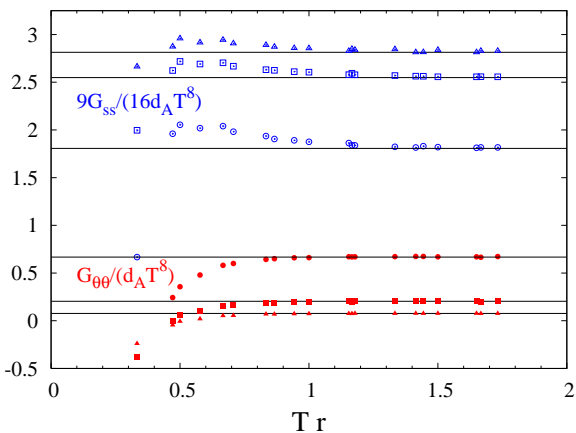,angle=-90,width=8.5cm}}
\caption{The $\Nt=6$ entropy-entropy ($G_{ss}$) and action-action 
($G_{\theta\theta}$)
correlators at $1.24$ (circles), $1.65$ (squares) and $2.20T_c$ (triangles).}
\la{fig:act-entr-1p24}
\end{figure}
%%%%%%%%%%%%%%%%%%%%%%%%%%%%%%%%%%%%%%%%%%%%%%%%%%%%%%%%%%%%%%%%%%%%%%%
\section*{Numerical results}
We now turn to a calculation of $G_{ee}(T,r)$ in the SU(3) gauge theory,
i.e. in the plasma of gluons, using lattice Monte-Carlo techniques on a
$(1/T)\cdot L^3$ lattice.
We employ the (isotropic) Wilson action~\cite{wilson74}
and the `once-HYP-smeared' `clover' discretization of $\theta_{00}$ and $\theta$
developed in~\cite{gluex}. The variance of $\<\theta_{00}\>$ was shown~\cite{Meyer:2007ed}
to be reduced by almost two orders of magnitude as compared to the simplest `plaquette'
discretization. This technical improvement allows us to obtain a signal 
for $G_{ee}-e^2(T)$ out to $r\simeq 1.5/T$.

% Figure \ref{fig:T0} shows the energy density correlator 
% at $T=0$, the quantity we will be subtracting from the 
% thermal correlator in the following. For $r/a\geq 3$
% comparison of the data obtained at three lattice spacings 
% shows that discretization errors are under control.

Figure \ref{fig:T0} displays the relevant correlators
at zero temperature. They fall off monotonically 
as $r^{-8}$ at short distance and exponentially at large distance.
Note that the trace anomaly correlator, while $O(\alpha_s^2)$ 
at short distances, dominates at large distances.
For $r/a\geq 3$
comparison of the data obtained at three lattice spacings 
shows that discretization errors are under control.

Figure \ref{fig:ee-b6p018} shows the 
qualitative change of $G_{ee}$ across the deconfining
phase transition. At $T=1.24T_c$ the functions obtained 
from these two lattice spacings are in qualitative agreement,
a non-trivial check, given the large cancellation between
the finite and zero temperature correlators. 
% The figure also illustrates how quickly the variance increases with 
% decreasing lattice spacing. 
The function $G_{ee}(r)$ is large and negative at short-distances, crosses
the asymptotic value $e^2(T)$, reaches a maximum and 
presumably decreases monotonically after that. 
Although the signal becomes too small
to tell beyond $Tr\simeq 1.5$, this is plausible 
in view of the small value of the thermal 
screening mass (see next section).

The $1/r^4$ short-distance divergence
is not visible below $T_c$, a fact that the OPE and the smallness
of $(e,p)$ in the confined phase easily account for.
We note that $G_{ee}/e^2(T)$  reaches 
around $r=0.6/T$ a maximum which is larger at $1.08$
than at $1.24T_c$.  We understand this in terms of the 
larger fluctuations present near the (weakly) first order
phase transition. 

Figure \ref{fig:ee-nt6} shows the 
temperature dependence of $G_{ee}$ up to $2.2T_c$.
The position of the maximum remains  $r_{\rm max}\approx0.6/T$,
and $G_{ee}(r_{\rm max},T)/e^2(T)$ decreases slowly 
as  the temperature rises. The curves at 1.65 and $2.20T_c$ exhibit
near-conformal behavior (i.e., $G_{ee}$ is essentially a function 
of $Tr$), however the asymptotic approach to $e^2(T)$ has 
the opposite sign, as a study of screening masses shows.

Figure \ref{fig:act-entr-1p24} shows separately the entropy 
density correlator and the trace anomaly correlator.
The former is qualitatively similar to the energy density
correlator, while the latter has a rather featureless monotonic 
behavior.

\section*{Screening masses}
The asymptotic large-$r$ behavior of $G_{ee}$ is dictated 
by the smallest screening mass that $T_{00}$ couples to.
This is the state invariant under all the symmetries of a
constant `$z$-slice'~\cite{karsch93,gupta98}, which has a volume
 $(1/T)\times L\times L$.
From $D=4$ simulations, we obtain directly
\be
% M(T) 
% 6.200  Nt=8   4.500000   0.32736104      0.00645   
% 6.200  Nt=6   4.500000   0.47075178      0.00943
% 6.408  Nt=6   2.500000   0.49989509      0.00424
%
% M(T=0):
%  6.408:  0.3919(92)
%  6.200:  0.5197(51)
\frac{M(T)}{M_4}=
 0.630(14),%  0.658(15),    
~~0.906(20),    % 0.927(18),  % 9263 meast
~~1.276(32)   % 
\ee
respectively at $1.24$ ($\Nt=8$), $1.65$, and $2.20T_c$ ($\Nt=6$).
So it is only at $T^*=1.790(36)T_c$  that 
the thermal screening starts to exceed the $D=4$ glueball mass.
In particular, for $T>T^*$, $G_{ee}(r,T)$ approaches its asymptotic value 
from below, and therefore crosses $e^2(T)$ twice.

% The velocity of sound can be obtained from 
% \be
% (\frac{1}{v_s^2}-3)\frac{s}{T^3}=
% \left(4-\frac{d^2\beta/d(\log a)^2}{d\beta/d\log a}\right)\frac{e-3p}{T^4}
% +\frac{1}{T^4}\<\sum_{x}\theta(x)\theta(0)\>_{T-0}
% \ee

\section*{Comparison with non-relativistic systems}
The radial distribution function $g(r)$ of a 
simple non-relativistic liquid, such as $^{36}$Ar at 85K~\cite{argon},
exhibits several very pronounced peaks above 1.
In particular $g(r)-1$ is of order unity at the first peak.
However, it is known~\cite{csernai} that $\eta/s$ is 
minimal near the liquid-gas phase transition, and becomes large
both at low and high temperature. A highly ordered
mesoscopic scale favours the transport of momentum,
because the holes between the closely packed molecules then
play the role of quasiparticles with a long mean free path
(an argument attributed to Enskog~\cite{csernai}).
Heating up the liquid has the effect of reducing the 
amplitude of the peaks in $g(r)$, until they disappear 
completely once the system is in a dilute gas phase.
Thus the regime where $\eta/s$ is minimal is the one 
where $g(r)$ has only few, small oscillations around unity.

We have found that $G_{ee}$ has exactly one 
broad peak above $e^2(T)$, exceeding that value by about $5-15\%$
(the figure decreases slowly with temperature). By analogy 
with non-relativistic fluids, it is tempting to see a 
relation between this fact and the small value obtained for
the shear viscosity~\cite{hm-shear} in the same range
of temperatures, $\eta/s<1.0$.

\section*{Conclusion}
We have calculated non-perturbatively the thermal part $G_{ee}$ 
of the energy-density 
spatial correlator in the plasma of SU(3) gluons, in the range of temperatures
$0.9\leq T/T_c\leq2.2$, and found the qualitative high-$T$ behavior. 
It diverges as $-r^{-4}$ near the origin as dictated by the OPE, 
and above $T_c$ reaches a maximum at $r_{\rm max}\approx0.6/T$ 
typically $10\%$ above its asymptotic value $e^2(T)$. 
For $T<1.8T_c$ it asymptotically approaches that value from above, 
and at higher temperatures it approaches from below. 
While the appearance of $G_{ee}$ is rather different 
from the  radial distribution function of a typical liquid,
we pointed out that it is precisely when the short-range order
is weak that the ratio $\eta/s$ is minimal.

% We note that at large $N_c$ the disconnected part $e^2(T)=O(N_c^4)$, 
% while the connected part of $G_{ee}$ is O($N_c^2$) at short distance 
% and O($N_c^0$) at long distance, so that the relative 
% size of the energy density fluctuations is 
% suppressed. Nevertheless the correlations in
% the thermal fluctuations
% are still large compared to the energy density of the confined phase.

We note that the radial distribution function of monopoles has been
computed in SU(2) gauge theory~\cite{D'Alessandro:2007su}, 
and has a similar shape to $G_{ee}$.
Color charge and monopole-antimonopole correlators 
that look alike have also been found in models~\cite{Gelman:2006xw}.
It would be interesting to see whether such models can reproduce $G_{ee}$.

% This does not however imply that the large-$N_c$ 
% QGP is a dilute gas immediately above $T_c$. These correlations in
% the thermal fluctuations
% are still large compared to the energy density of the confined phase.
% and can trigger the transition to confinement through bubble growth.

A straightforward extension of this work is to 
consider the full spatial correlations of the energy momentum tensor
$\<T_{\mu\nu}T_{\rho\sigma}\>$, in SU(3) gauge theory and in full QCD. 
As a benchmark it would be helpful 
to know the form of these correlators in the strongly coupled 
${\cal N}=4$ SYM theory, which is known to be an excellent fluid
from the smallness of $\eta/s$~\cite{policastro} and from
its spectral functions~\cite{Kovtun:2006pf}.

I thank Hong Liu, John Negele and Krishna Rajagopal for 
their encouragement and stimulating discussions. I further thank Derek~Teaney 
for discussions at the ``QCD under Extreme Conditions'' Workshop
(North Carolina State University, 21-23 July 2008).
The simulations were done on the desktop machines of the 
Laboratory for Nuclear Science at M.I.T. This work was supported in part by 
funds provided by the U.S. Department of Energy under cooperative research agreement
DE-FG02-94ER40818.

%%%%%%%%%%%%%%%%%%%%%%%%%%%%%%%%%%%%%%%%%%%%%%%%%%%%%%%%%%%%%%%%%%%%%%%%%%%

\end{document}